\documentclass[aps,prl, reprint, 
showpacs,showkeys,superscriptaddress,preprintnumbers,nofootinbib]{revtex4-1} 
\usepackage{amsmath,amssymb,bm,mathrsfs}
\usepackage{hyperref}
\usepackage{url}
\usepackage{graphicx}
\usepackage{color}

\begin{document}

\preprint{\tt STUPP-21-249}

\title{Probing the $L_\mu - L_\tau$ Gauge Boson at the MUonE Experiment} 

\author{Kento Asai}
\email{asai@krishna.th.phy.saitama-u.ac.jp}
\affiliation{Department of Physics, Faculty of Science, Saitama University, Sakura-ku, Saitama 338--8570, Japan}

\author{Koichi Hamaguchi}
\email{hama@hep-th.phys.s.u-tokyo.ac.jp}
\affiliation{Department of Physics, University of Tokyo, 
Tokyo 113--0033,
Japan}
\affiliation{
Kavli Institute for the Physics and Mathematics of the Universe (Kavli
IPMU), University of Tokyo, Kashiwa 277--8583, Japan
}

\author{Natsumi Nagata}
\email{natsumi@hep-th.phys.s.u-tokyo.ac.jp}
\affiliation{Department of Physics, University of Tokyo, 
Tokyo 113--0033,
Japan}

\author{Shih-Yen Tseng}
\email{shihyen@hep-th.phys.s.u-tokyo.ac.jp}
\affiliation{Department of Physics, University of Tokyo, 
Tokyo 113--0033,
Japan}

\author{Juntaro Wada}
\email{wada@hep-th.phys.s.u-tokyo.ac.jp}
\affiliation{Department of Physics, University of Tokyo, 
Tokyo 113--0033,
Japan}

\begin{abstract}

We discuss the prospects of probing the $L_\mu - L_\tau$ gauge boson at the MUonE experiment. The $L_\mu - L_\tau$ gauge boson $Z^\prime$ with a mass of $\lesssim 200$~MeV, which can explain the discrepancy between the measured value of the muon $g-2$ and the value calculated in the Standard Model, can be produced at the MUonE experiment through the process $\mu e \to \mu e Z^\prime$. The $Z^\prime$ in the final state decays into a pair of neutrinos, and therefore we cannot observe the decay of $Z^\prime$ directly. It is, however, still possible to probe this signature by searching for events with a large scattering angle of muon and a less energetic final-state electron. The background events coming from the elastic scattering $\mu e \to \mu e$ as well as radiative process $\mu e \to \mu e \gamma$ can be removed by the kinematical cuts on the muon scattering angle and the electron energy, in addition to a photon veto. The background events from the electroweak process $\mu e \to \mu e \nu \bar{\nu}$ are negligible. With our selection criteria, the number of signal events $\mu e \to \mu e Z^\prime$ is found to be as large as $\sim 10^3$, assuming an integrated luminosity of $15~\mathrm{fb}^{-1}$, in the parameter region motivated by the muon $g-2$ discrepancy. It is, therefore, quite feasible to probe the $L_\mu - L_\tau$ gauge boson at the MUonE experiment---without introducing additional devices---and we strongly recommend recording the events relevant to this $Z^\prime$ production process. 
%Elastic scattering events, $\mu e \to \mu e$, do not yield such a signature as a large muon scattering angle is always accompanied by an energetic electron.  The $\mu e \to \mu e \gamma$ process, on the other hand, can give rise to this signature only when there is a hard photon, and thus we can remove this process by requiring a photon veto. The processes including neutrinos, $\mu e \to \mu e \nu \bar{\nu}$, do yield the same signature as the $\mu e \to \mu e Z^\prime$ process; these processes are induced via the exchange of the electroweak gauge bosons in the Standard Model. We, however, find that the number of events of these electroweak processes is negligibly small, while that of $\mu e \to \mu e Z^\prime$ is as large as $\sim 10^3$ in the parameter region motivated by the muon $g-2$ discrepancy. It is, therefore, quite feasible to probe the $L_\mu - L_\tau$ gauge boson at the MUonE experiment, without introducing additional devices, and we strongly recommend to record the events relevant to this $Z^\prime$ production process. 

\end{abstract}

\maketitle

%%%%%%%%%%%%%%%%%%%%%%%
\section{Introduction}
%%%%%%%%%%%%%%%%%%%%%%%

The latest measurement of the anomalous magnetic moment ($g-2$) of muon by the Fermilab Muon $g-2$ Experiment~\cite{Muong-2:2021ojo}, combined with the previous result by the BNL E821 experiment~\cite{Muong-2:2006rrc}, shows that the measured value of the muon $g-2$ deviates from the value calculated in the Standard Model (SM)~\cite{Aoyama:2020ynm} by 4.2$\sigma$. The largest uncertainty in the SM calculation at the present moment comes from the hadronic vacuum polarization (HVP) contribution. In Ref.~\cite{Aoyama:2020ynm}, the Muon $g-2$ Theory Initiative determined the HVP contribution from $e^+ e^-$ data~\cite{davier:2017zfy, keshavarzi:2018mgv, colangelo:2018mtw, hoferichter:2019gzf, davier:2019can, keshavarzi:2019abf, kurz:2014wya}. A recent lattice QCD simulation~\cite{Borsanyi:2020mff}, however, found a value of the HVP contribution larger than that presented in Ref.~\cite{Aoyama:2020ynm}, which considerably relaxes the $g-2$ discrepancy. It is, thus, of great importance to improve the determination of the HVP contribution in order to clarify the situation. 

The MUonE experiment~\cite{Abbiendi:2016xup, Abbiendi:2677471} aims at determining the HVP contribution with a method~\cite{CarloniCalame:2015obs} different from the aforementioned ones. In this experiment, muons with an energy of 150~GeV collide with electrons at rest. Through precise measurements of the differential scattering cross sections of the $\mu e \to \mu e$ process, the size of the HVP contribution in space-like momentum region is extracted, from which one can determine the HVP contribution to the muon $g-2$. The precision of this evaluation is expected to be comparable to or smaller than the present uncertainty in the calculation of the HVP contribution and, in particular, is smaller than the size of the $g-2$ discrepancy by about an order of magnitude. A test run of the MUonE experiment is scheduled in 2021~\cite{Abbiendi:2020sxw}; if it is successful, full running may be performed in 2022--24~\cite{Venanzoni}. 

In the meantime, there have been a variety of proposals to explain the muon $g-2$ discrepancy in models beyond the SM. A simple, successful class of models are based on the $L_\mu - L_\tau$ gauge theory~\cite{Foot:1990mn, He:1990pn, He:1991qd, Foot:1994vd}, where a massive gauge boson associated with this gauge symmetry, called the $L_\mu - L_\tau$ gauge boson ($Z^\prime$), contributes to the muon magnetic moment at one-loop level. This type of gauge theory was first considered as a potential way to promote a global symmetry in the SM to a gauge symmetry. As it turns out, this gauge theory allows the introduction of right-handed neutrinos as well. Moreover, a minimal setup with three right-handed neutrinos is found to provide a neutrino-mass structure compatible with the current neutrino experimental data~\cite{Asai:2017ryy, Asai:2018ocx, Asai:2019ciz, Asai:2020qax}, even though it is highly constrained by this gauge symmetry due to its flavor dependence~\cite{Branco:1988ex, Choubey:2004hn, Araki:2012ip, Heeck:2014sna, Crivellin:2015lwa, Plestid:2016esp}. As discussed in a number of previous studies~\cite{Baek:2001kca, Ma:2001md, Heeck:2011wj, Harigaya:2013twa, Altmannshofer:2014pba, Araki:2014ona, Kamada:2015era, Araki:2015mya, Baek:2015fea, Fuyuto:2015gmk, Patra:2016shz, Biswas:2016yan, Ibe:2016dir, Biswas:2016yjr, Kaneta:2016uyt, Araki:2017wyg, Chen:2017cic, Gninenko:2018tlp, Nomura:2018yej, Bauer:2018onh, Kamada:2018zxi, Banerjee:2018eaf,
Crivellin:2018qmi,Foldenauer:2018zrz, Banerjee:2018mnw, Escudero:2019gzq, Altmannshofer:2019zhy, Krnjaic:2019rsv, Ballett:2019xoj, Biswas:2019twf, Amaral:2020tga, Borah:2020jzi, Shimomura:2020tmg, Asai:2020qlp,Banerjee:2020zvi, Zhang:2020fiu, Huang:2021nkl, Araki:2021xdk, Banerjee:2021laz, Amaral:2021rzw, Zu:2021odn, Borah:2021jzu,  Zhou:2021vnf, Carpio:2021jhu,
Qi:2021rhh, Borah:2021mri, Greljo:2021npi, Holst:2021lzm, Drees:2021rsg, Hapitas:2021ilr, Borah:2021khc}, the observed value of the muon $g-2$ discrepancy can be explained by the $L_\mu - L_\tau$ gauge models while evading the current experimental limits, for a mass of $Z^\prime$ in the range $m_{Z^\prime}\sim 10$--200~MeV .

As the MUonE experiment is designed to determine the HVP contribution with a precision much better than the 4.2$\sigma$ discrepancy, one may expect that it is also sensitive to the new physics contributions accounting for the discrepancy. This expectation, however, turns out to be generically incorrect---as shown in Refs.~\cite{Schubert:2019nwm, Dev:2020drf, Masiero:2020vxk}, the measurement of the $\mu e \to \mu e$ elastic scattering at the MUonE experiment is actually insensitive to new physics effects. In particular, the contribution of $Z^\prime$ in the $L_\mu - L_\tau$ models to the $\mu e \to \mu e$ process, which is induced at one-loop level, is smaller than the HVP contribution by orders of magnitude, and the parameter regions where this measurement has sensitivities have already been excluded by other experiments. 

%Nevertheless, we point out in this paper that it is in fact possible to probe the $L_\mu - L_\tau$ models at the MUonE experiment, by searching for the signature associated with the direct production of $Z^\prime$, $\mu e \to \mu e Z^\prime$. 

Nevertheless, we point out in this letter that we can probe the $L_\mu - L_\tau$ models at the MUonE experiment by searching for the signature associated with the direct production of $Z^\prime$, $\mu e \to \mu e Z^\prime$.

%%%%%%%%%%%%%%%%%%%%%%%
\section{Kinematics}
%%%%%%%%%%%%%%%%%%%%%%%

The MUonE experiment~\cite{Abbiendi:2016xup, Abbiendi:2677471} plans to use the 150~GeV muon beam at the CERN North Area, with the target being electrons in beryllium atoms. The experimental apparatus consists of a series of 40 stations, each of which has a 15~mm thick Be target and tracking sensors. Right after these 40 stations, an electromagnetic calorimeter (ECAL) is located. A muon filter with muon chambers is placed at the end. See Letter of Intent~\cite{Abbiendi:2677471} for more details on the detector setup and its projected performance. 

The primary target of the MUonE experiment is the elastic scattering process, $\mu e \to \mu e$, where the initial-state electron is at rest. Given the initial energy of muon $E_{\mu, i} = 150~\mathrm{GeV}$, the energies and the scattering angles of the final-state electron and muon are determined as functions of one parameter. For example, the muon scattering angle $\theta_\mu$ is related to the electron scattering angle $\theta_e$ by 
\begin{equation}
  \tan \theta_\mu = \frac{2 \tan \theta_e}{(1 + R) (1 + \gamma^2 \tan^2 \theta_e) - 2} ~,
  \label{eq:tantmu}
\end{equation}
where $R \equiv (m_\mu^2 + m_e E_{\mu, i})/(m_e^2 + m_e E_{\mu, i})$ with $m_\mu$ and $m_e$ the masses of muon and electron, respectively, and $\gamma \equiv (m_e + E_{\mu, i})/\sqrt{s}$ with $\sqrt{s}$ the center-of-mass energy: 
\begin{equation}
  s = m_\mu^2 + m_e^2 + 2 m_e E_{\mu,i} ~.
  \label{eq:s}
\end{equation}
In addition, the final-state electron energy $E_e$ is given by 
\begin{equation}
  E_e = m_e \frac{1 + \beta^2 \cos^2 \theta_e}{1 - \beta^2 \cos^2 \theta_e} ~,
  \label{eq:ee}
\end{equation}
with $\beta \equiv \sqrt{E_{\mu,i}^2 - m_\mu^2}/(E_{\mu, i} + m_e)$. From Eq.~\eqref{eq:tantmu}, we see that there is a maximum value of $\theta_\mu$, $\theta_\mu^{(\mathrm{max})} = 1/(\gamma \sqrt{R^2-1}) \simeq 4.84~\mathrm{mrad}$. 

%%%%%%%%%%%%%%%%%%%%%%%%%%%%%%%%%%%%%%%%%%%%%%%%%%%%%%%%%%%%%%%%%%%%
\begin{figure}[t]
  \centering
  {\includegraphics[width=0.45\textwidth]{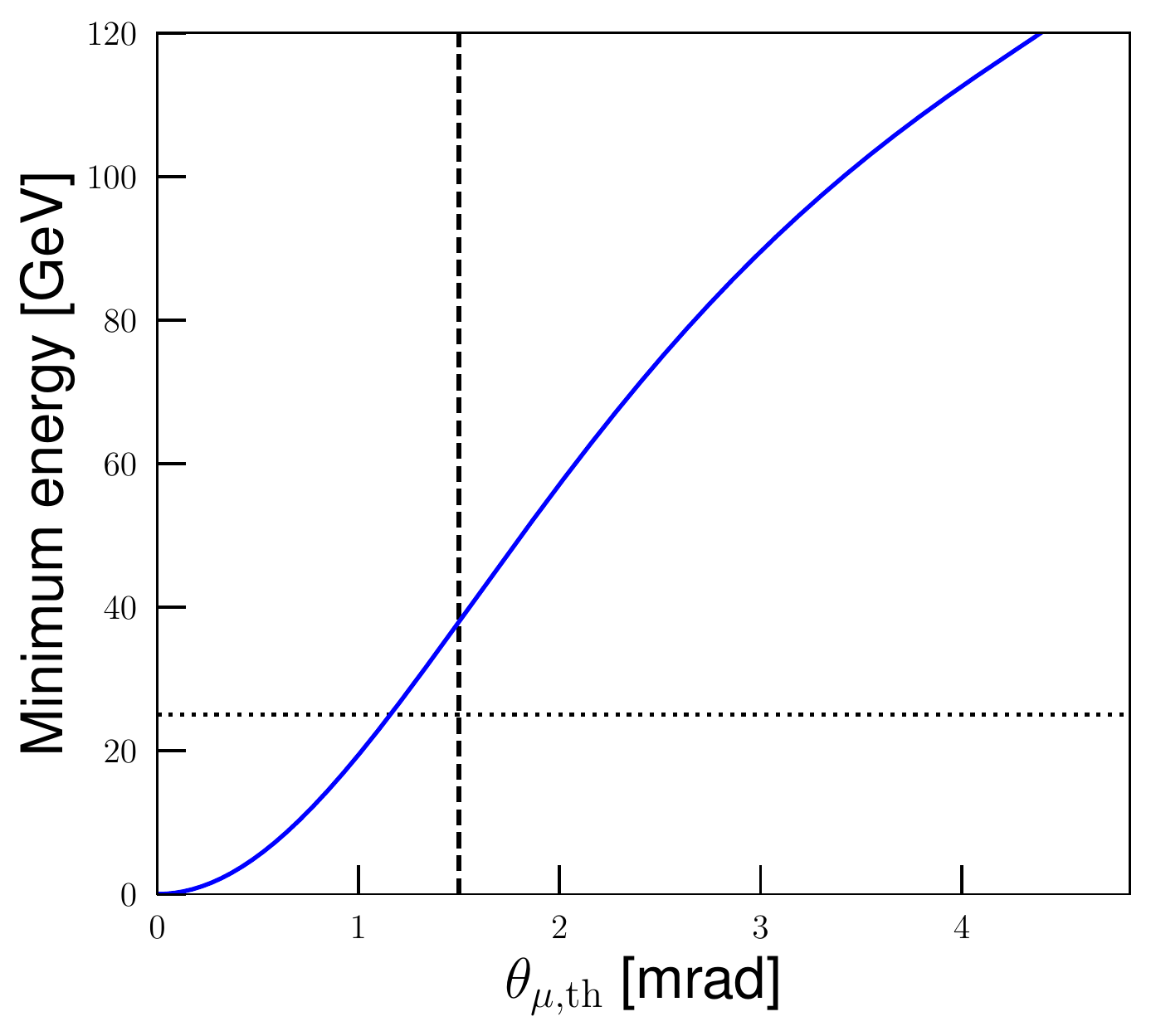}} 
  \caption{
      Minimum value of $E_e$ ($E_{e\gamma}$) under the condition $\theta_\mu > \theta_{\mu, \mathrm{th}}$ as a function of $\theta_{\mu, \mathrm{th}}$ for $\mu e \to \mu e$ ($\mu e \to \mu e \gamma$). The vertical dashed and horizontal dotted lines correspond to the threshold values of $\theta_\mu$ and $E_e$ we require for our selection criteria, respectively. 
  }
  \label{fig:eeminvsthmu}
\end{figure}
%%%%%%%%%%%%%%%%%%%%%%%%%%%%%%%%%%%%%%%%%%%%%%%%%%%%%%%%%%%%%%%%%%%%%

Suppose that we require $\theta_\mu$ to be larger than a certain threshold value $\theta_{\mu, \mathrm{th}} < \theta_\mu^{(\mathrm{max})}$. This restricts $\theta_e$, and thus $E_e$ as well, to be in a finite range, according to Eq.~\eqref{eq:tantmu} and Eq.~\eqref{eq:ee}, respectively. In particular, the maximum value of $\theta_e$ within this range corresponds to the minimum value of $E_e$ for $\theta_\mu > \theta_{\mu, \mathrm{th}}$, which we denote by $E_e^{(\mathrm{min})}$; in Fig.~\ref{fig:eeminvsthmu}, we plot $E_e^{(\mathrm{min})}$ as a function of $\theta_{\mu, \mathrm{th}}$ . 

Another process that could contaminate the elastic scattering process is the $\mu e \to \mu e \gamma$ process. For later use, we consider the minimum value of $E_{e\gamma} \equiv E_e + E_\gamma$ under the condition $\theta_\mu > \theta_{\mu, \mathrm{th}}$, where $E_\gamma$ is the energy of the final-state photon. To obtain this, we define $m_{e\gamma}^2 \equiv (p_e + p_\gamma)^2$, where $p_e$ and $p_\gamma$ are the four-momenta of the final-state electron and photon, respectively. In the center-of-mass frame, this quantity is evaluated as $m_{e\gamma}^2 = s + m_\mu^2 - 2 \sqrt{s} E_{\mu, \mathrm{cm}}$, with $E_{\mu, \mathrm{cm}}$ the out-going muon energy in this frame. We see that $E_{\mu, \mathrm{cm}}$, and therefore $E_\mu$ as well, is maximized for a minimum value of $m_{e\gamma}^2$. This consequence does not depend on the muon scattering angle. Because of the energy-conservation law, $E_{e\gamma}$ is minimized when $E_\mu$ is maximized; hence, the minimum value of $E_{e\gamma}$ is obtained when $m_{e\gamma}^2$ is minimized. Now note that $m_{e\gamma}^2 > m_e^2$, and $m_{e\gamma}^2 \to m_e^2$ for $E_\gamma \to 0$, \textit{i.e.}, in the soft photon limit. Since this limit corresponds to the elastic scattering discussed above, we conclude that $E_{e\gamma} > E_e^{(\mathrm{min})}$ for $\theta_\mu > \theta_{\mu, \mathrm{th}}$, and thus Fig.~\ref{fig:eeminvsthmu} can also be regarded as the lower limit on $E_{e\gamma}$ in the $\mu e \to \mu e \gamma$ process under the condition $\theta_\mu > \theta_{\mu, \mathrm{th}}$.

From Eq.~\eqref{eq:s}, we have $\sqrt{s} \simeq 406$~MeV. This means that we can create a new particle at the MUonE if its mass is $\lesssim 300$~MeV. In particular, the $L_\mu - L_\tau$ gauge boson with a mass $m_{Z^\prime} \lesssim 300$~MeV can be produced through the process $\mu e \to \mu e Z^\prime$. The $Z^\prime$ decays into a pair of neutrinos, and therefore cannot be observed directly. We, however, show below that it is still possible to detect this signature if we impose appropriate selection conditions.

%%%%%%%%%%%%%%%%%%%%%%%%%%%%%%%%%%%%%%%%
\section{Search strategy and results}
%%%%%%%%%%%%%%%%%%%%%%%%%%%%%%%%%%%%%%%%

Let us now describe the search strategy and show its prospects for the $Z^\prime$ analysis. To remove the SM background processes, we impose the following selection criteria: 
\begin{itemize}
  \item[(i)] $\theta_\mu > 1.5$~mrad. 
  \item[(ii)] $1~\mathrm{GeV} < E_e < 25~\mathrm{GeV}$. 
  \item[(iii)] Photon veto.
\end{itemize}
Note that $\theta_\mu$ can be measured with a resolution of $\mathcal{O} (0.01)$~mrad at the MUonE. $E_e$ and $E_\gamma$ are measured by the ECAL, which is expected to offer an energy resolution of $\lesssim 10$\%.
\footnote{Our strategy relies on the resolution of the ECAL, and the sensitivity of the muon angle, but is independent of the electron angle which tends to be affected by the multiple scattering.}

We show the minimum value of $\theta_\mu$ in (i) and the maximum value of $E_e$ in (ii) in the vertical dashed and horizontal dotted lines in Fig.~\ref{fig:eeminvsthmu}, respectively. From this plot, it can be immediately seen that the elastic scattering process never occurs under the conditions (i) and (ii); with (i), the minimum value of $E_e^{(\mathrm{min})}$ is $\simeq 38$~GeV, which is out of the range of (ii).\footnote{In actual experiments, a small fraction of the elastic scattering events may leak into the signal region due to the energy resolution of the ECAL, which is estimated to be $\sim 2$--5\% for $E_e \gtrsim 40$~GeV depending on the position at which the electron is produced~\cite{Abbiendi:2677471}. This position dependence originates from the secondary interactions with the silicon detectors and Be targets, and by reconstructing the pattern of hits along the electron track it may be possible to improve the energy resolution~\cite{Abbiendi:2677471}. In any case, the conditions (i) and (ii) should be tuned after the detector calibration so that the contamination of the elastic scattering events is suppressed sufficiently.   } 

The $\mu e \to \mu e \gamma$ process can occur under the conditions (i) and (ii), but in this case $E_\gamma$ must be larger than $38~\mathrm{GeV} - E_e \gtrsim 13$~GeV. Such an energetic photon can be detected at the ECAL if the final-state electron and photon are well separated,\footnote{The angular resolution of the ECAL is $\lesssim 1$~mrad~\cite{Abbiendi:2677471}. } and thus would be vetoed by the condition (iii). We can show that the photon is emitted in the forward direction, $\theta_\gamma \lesssim 6$~mrad;\footnote{
To see this, note that in the $\mu e \to \mu e \gamma$ process, the sum of the final-state electron and photon momenta, $\bm{p}_{e\gamma} \equiv \bm{p}_e + \bm{p}_\gamma$, is fully determined as a function of $E_\mu$ and $\theta_\mu$, and so is its angle with respect to the beam axis, $\theta_{e\gamma}$. The angle between $\bm{p}_{\gamma}$ and $\bm{p}_{e\gamma}$, $\theta_{e\gamma, \gamma}$, is given by 
\begin{equation}
    \cos \theta_{e\gamma, \gamma} = \frac{E_{e\gamma}}{|\bm{p}_{e\gamma}|} - \frac{m_{e\gamma}^2 - m_e^2}{2 |\bm{p}_{e\gamma}|(E_{e\gamma}-E_e)} ~.
    %E_\gamma} ~.
\end{equation}
As noted above, $m_{e\gamma}^2 > m_e^2$, and thus $\theta_{e\gamma, \gamma}$ is maximized when 
%$E_\gamma$ is minimized, \textit{i.e.}, for $E_\gamma \simeq 13$~GeV. 
$E_e$ is maximized, \textit{i.e.}, for $E_e \simeq 25$~GeV. 
This maximum value, $\theta_{e\gamma, \gamma}^{(\text{max})}$, sets an upper limit on $\theta_\gamma$ for a given set of $E_\mu$ and $\theta_\mu$: $\theta_\gamma < \theta_{e\gamma, \gamma}^{(\text{max})} + \theta_{e\gamma}$. By varying $E_\mu$ and $\theta_\mu$ within their allowed range, we then obtain an upper limit on $\theta_\gamma$: $\theta_\gamma \lesssim 6$~mrad. 
} the present design of the MUonE experiment supposes the ECAL transverse dimension of ${\cal O}(1 \times 1)~{\rm m}^2$~\cite{Abbiendi:2677471}, with which all the photons are expected to hit the ECAL.
%\footnote{For a smaller size of the ECAL, we should include only the events that occur at positions within the photon coverage or tune the selection criteria such that all of the photons must hit the ECAL. In these cases, the expected number of signal events decreases according to the coverage fraction or due to the tighter selection criteria. Another option is to introduce additional gamma-ray detectors for the photon veto to achieve full coverage.
%As we see below, we expect $\sim 10^3$ signal events in the parameter regions of our interest, and therefore even if we cover the regions only within ${\cal O}(1)$~m from the end of the stations, we still expect a sizable number of signal events.  
%} 
%To avoid this problem, we propose to cover the stations with gamma-ray detectors used primarily for the photon veto.
If, on the other hand, the electron and photon cannot be distinguished in the ECAL, they are detected as a single electron having energy larger than 38~GeV---this event is eliminated by the condition (ii). Therefore, both the $\mu e \to \mu e $ and $\mu e \to \mu e \gamma$ processes can be safely removed with our selection criteria, just by kinematics. 

Other potential background sources include multiple scattering associated with the $\mu e \to \mu e$ process, muon-nuclear scattering, and $\mu e \to \mu e \nu \bar{\nu}$ induced by the electroweak gauge boson exchange. The significance of the first two strongly depends on the experimental setup, which is not fully fixed yet; hence, the evaluation of the number of events associated with these background sources is beyond the scope of this letter. We, however, expect that these two can be well controlled. Multiple scattering may mimic the signal-like signature if the muon scattering angle satisfies the condition (i) and the electron loses its energy by a subsequent scattering to fall into the signal range (ii); or the final-state electron energy is within the range (ii) and the muon undergoes a second scattering so that its scattering angle gets larger to satisfy (i). In both cases, there should be a sizable energy deposit at the second scattering point, which can be detected as a kink/branch of the track. For Muon-nuclear scattering events, they are expected to be identified by using track multiplicity as discussed in Ref.~\cite{Abbiendi:2677471}.
%Muon-nuclear scattering events are, on the other hand, expected to %be identified by using track multiplicity as discussed in %Ref.~\cite{Abbiendi:2677471}.
On the contrary, the electroweak processes for $\mu e \to \mu e \nu \bar{\nu}$ cannot be eliminated by the kinematical cuts, since they yield the same final state as that of $\mu e \to \mu e Z^\prime$. We thus compute the number of these events using Monte Carlo simulations. 

%%%%%%%%%%%%%%%%%%%%%%%%%%%%%%%%%%%%%%%%%%%%%%%%%%%%%%%%%%%%%%%%%%%%
\begin{figure}[t]
  \centering
  {\includegraphics[width=0.47\textwidth]{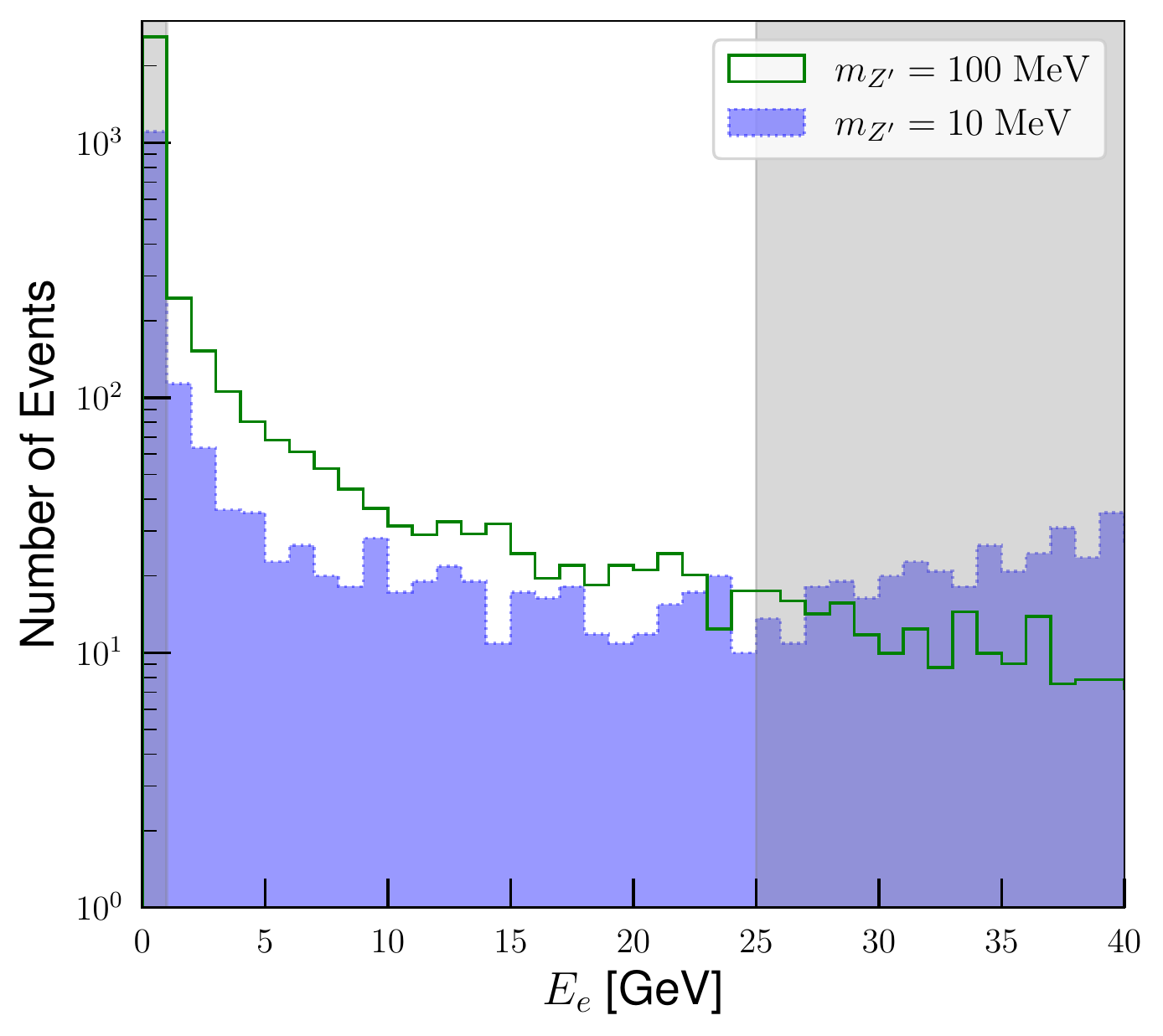}} 
  \caption{Distribution of $E_e$ of signal events for the two choices of the parameter points under the condition (i) for an integrated luminosity of $15~\mathrm{fb}^{-1}$. Gray shaded regions are out of the range of the condition (ii). 
  }
  \label{fig:ee_sig}
\end{figure}
%%%%%%%%%%%%%%%%%%%%%%%%%%%%%%%%%%%%%%%%%%%%%%%%%%%%%%%%%%%%%%%%%%%%%

Fig.~\ref{fig:ee_sig} shows the distribution of $E_e$ of signal events under the condition (i) for two sets of parameters: $m_{Z^\prime } = 100$~MeV and $g_{Z^\prime} = 10^{-3}$ (green); and $m_{Z^\prime } = 10$~MeV and $g_{Z^\prime} = 5 \times 10^{-4}$ (blue).
%, with $g_{Z^\prime}$ the gauge coupling of the $L_\mu - L_\tau$ gauge boson. 
The gauge coupling $g_{Z^\prime}$ is defined by the Lagrangian ${\cal L}_{\rm int}=-g_{Z^\prime}Z'_\mu \sum_\psi Q_\psi \bar{\psi}\gamma^\mu \psi$, where $Z'_\mu$ is the $L_\mu - L_\tau$ gauge boson and $Q_\psi=1$ ($-1$) for $\psi$ being the second (third) generation leptons.
These parameter points can explain the observed muon $g-2$ discrepancy, as shown below. To simulate the signal events, we use \texttt{FeynRules v2.3.48}~\cite{Christensen:2008py, Alloul:2013bka} to generate the UFO file for our model and \texttt{MadGraph5\_aMC@NLO v3.1.1}~\cite{Alwall:2014hca} for Monte Carlo simulations. Here we assume an integrated luminosity of $15~\mathrm{fb}^{-1}$, which is expected to be reached in a few years of data taking~\cite{Abbiendi:2677471}. As we see, the final-state electrons tend to be softer for a larger $Z^\prime$ mass. In both cases shown in this figure, a large number of events remain after further imposing the condition (ii), which is shown in the gray shade: $\simeq 600$ ($1200$) events for $m_{Z^\prime} = 10$~MeV (100~MeV)\footnote{The low-energy electrons produced at the upstream targets may not arrive at the calorimeter. If we discard the electrons with energy lower than $3$~($5$)~$\mathrm{GeV}$, the number of signals is reduced at most by a factor of $0.7$~($0.5$), but our main conclusion does not change.}. We also find that the electrons are scattered dominantly into the forward region ($\theta_e \lesssim 20$~mrad) and thus most of them are expected to hit the ECAL.\footnote{The reduction of the number of signal events due to the imperfect coverage of the ECAL, which is expected to be at most an ${\cal O}(1)$ factor for the planned size of the ECAL, should be evaluated once the detector setup is fixed.} On the other hand, the number of events for the electroweak processes $\mu e \to \mu e \nu \bar{\nu}$, which is also computed with \texttt{MadGraph5\_aMC@NLO v3.1.1}, is found to be negligibly small: $ \sim 10^{-4}$ events for the same integrated luminosity, $15~\mathrm{fb}^{-1}$.

%%%%%%%%%%%%%%%%%%%%%%%%%%%%%%%%%%%%%%%%%%%%%%%%%%%%%%%%%%%%%%%%%%%%
\begin{figure}[t]
  \centering
  {\includegraphics[width=0.47\textwidth]{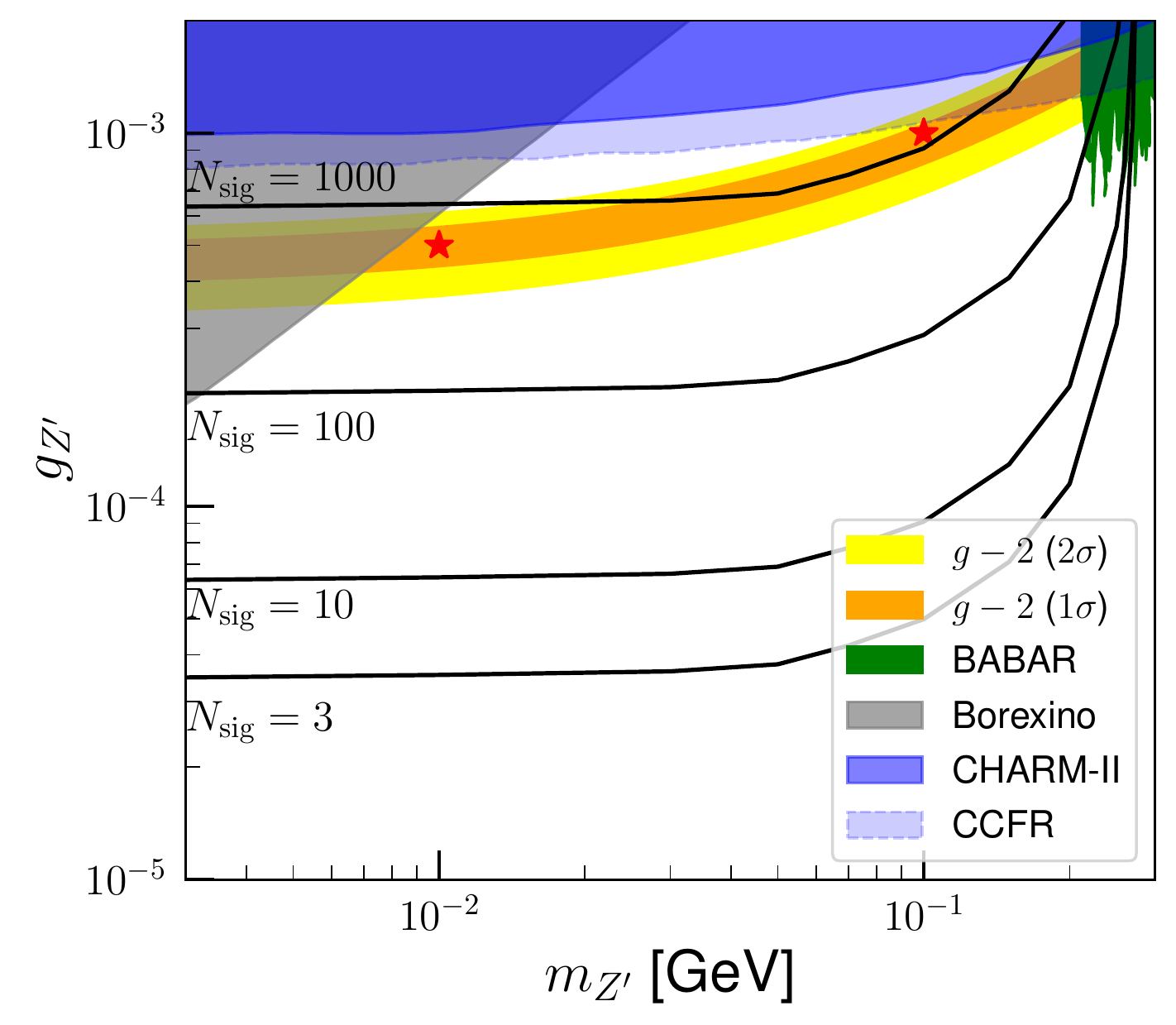}} 
  \caption{Contour of the number of signal events for an integrated luminosity of $15~\mathrm{fb}^{-1}$, $N_{\mathrm{sig}}$, under the conditions (i--iii) in the $m_{Z^\prime}$--$g_{Z^\prime}$ plane. The orange and yellow bands show the muon $g-2$ favored regions at the 1$\sigma$ and 2$\sigma$ levels, respectively. The two red stars are the benchmark points used in Fig.~\ref{fig:ee_sig}. The green, gray, dark blue, and light blue shaded areas are excluded by BABAR~\cite{BaBar:2016sci}, Borexino~\cite{Bellini:2011rx}, CHARM-II~\cite{CHARM-II:1990dvf}, and CCFR~\cite{CCFR:1991lpl}, respectively.
  }
  \label{fig:mzvsg}
\end{figure}
%%%%%%%%%%%%%%%%%%%%%%%%%%%%%%%%%%%%%%%%%%%%%%%%%%%%%%%%%%%%%%%%%%%%%

To show the potential impact of our search strategy, in Fig.~\ref{fig:mzvsg}, we show the contours of the number of signal events for an integrated luminosity of $15~\mathrm{fb}^{-1}$, $N_{\mathrm{sig}}$, under the conditions (i--iii) in the $m_{Z^\prime}$--$g_{Z^\prime}$ plane. The orange and yellow bands correspond to the parameter regions where the observed value of the muon $g-2$ discrepancy can be explained at the 1$\sigma$ and 2$\sigma$ levels, respectively. The two red stars are the benchmark points used in Fig.~\ref{fig:ee_sig}, which are found to be within the 1$\sigma$ band. The green-shaded area is excluded by the BABAR experiment~\cite{BaBar:2016sci}. The Belle experiment gives a similar limit in this region~\cite{Czank:2021nns}.   The gray-shaded region is disfavored by the neutrino-electron scattering data obtained at the Borexino experiment~\cite{Bellini:2011rx}; we take this bound from the result given in Ref.~\cite{Kamada:2018zxi}. The blue dark (light) shaded region represents the limit from the neutrino trident production processes imposed by the CHARM-II~\cite{CHARM-II:1990dvf} (CCFR~\cite{CCFR:1991lpl}) experiment, taken from Ref.~\cite{Altmannshofer:2014pba}. As seen in this figure, we expect $\sim 10^3$ signal events in the muon $g-2$ favored region. For low-mass regions, $\mathcal{O}(1)$ events are obtained for $g_{Z^\prime}$ as small as $\text{a few} \times 10^{-5}$. 
The sensitivity is comparable to the reach of NA64$\mu$~\cite{Sieber:2021fue} for $m_{Z^\prime} \lesssim 200$~MeV.
For $m_{Z^\prime} \gtrsim 100$~MeV, the number of events is suppressed kinematically, and vanishes at $m_{Z^\prime} \simeq 300$~MeV.

%%%%%%%%%%%%%%%%%%%%%%%
\section{Discussions}
%%%%%%%%%%%%%%%%%%%%%%%

We have discussed the prospects of probing the $L_\mu - L_\tau$ gauge boson at the MUonE experiment, by searching for events with (i) $\theta_\mu > 1.5$~mrad and (ii) $1~\mathrm{GeV} < E_e < 25~\mathrm{GeV}$. For the SM background, elastic scattering events, $\mu e \to \mu e$, do not yield such a signature as a large muon scattering angle is always accompanied by an energetic electron.  The $\mu e \to \mu e \gamma$ process can give rise to this signature only when a hard photon exists, and thus can be removed by requiring (iii) a photon veto. The number of events of the electroweak process $\mu e \to \mu e \nu \bar{\nu}$ turns out to be negligibly small. We find that the number of signal events $\mu e \to \mu e Z^\prime$ is as large as $\sim 10^3$ in the parameter region motivated by the muon $g-2$ discrepancy. It is, therefore, quite feasible to probe the $L_\mu - L_\tau$ gauge boson at the MUonE experiment, without introducing additional devices, 
%by introducing only gamma-ray detectors for photon veto, 
and we strongly recommend recording the events relevant to this $Z^\prime$ production process.
%, although they are regarded as ``background'' in the MUonE experiment. 

It is possible to improve our selection criteria (i--iii), by optimizing the threshold values and using additional variables, such as $\theta_e$, acoplanarity, etc. Such an optimization can be considered once the detector setup is fully fixed and its performance is well understood. Moreover, an additional detector to measure the muon energy/momentum could be useful to improve the search strategy; with this, we can easily detect the missing-energy carried by $Z^\prime$. We also note that with such a detector, it is in principle possible to fully reconstruct the four momentum of the produced $Z^\prime$ and, in particular, to measure its mass. The size of $g_{Z^\prime}$ can also be estimated from the number of events. This information allows us to test the explanation of the muon $g-2$ discrepancy with the $L_\mu - L_\tau$ gauge models. 

There are other potential SM background processes not quantitatively discussed in this work, due to the lack of knowledge on the actual experimental setup and detector performance. Although we expect they are controllable as argued above, the validation of this argument and precise evaluation of the number of events associated with the processes are certainly required. We will return to these and other issues in the future.

%%%%%%%%%%%%%%%%%% Acknowledgements %%%%%%%%%%%%%%%%%%%%%%%%%%%%%%%%%%%%%%%
\section*{Acknowledgments}
This work is supported in part by the Grant-in-Aid for Innovative Areas (No.19H05810 [KH], No.19H05802 [KH], No.18H05542 [NN]), Scientific Research B (No.20H01897 [KH and NN]), Young Scientists (No.21K13916 [NN]), Research Activity Start-up (No.21K20365 [KA]), and JSPS KAKENHI Grant (No.20J22214 [SYT]).
%%%%%%%%%%%%%%%%%%%%%%%%%%%%%%%%%%%%%%%%%%%%%%%%%%%%%%%%%%%%%%%%%%%%%%%%%%%

%%%%%%%%%%%%%%%%% Ref %%%%%%%%%%%%%%%%%%%%%%%%
\bibliographystyle{utphysmod}
\bibliography{ref}
%%%%%%%%%%%%%%%%%%%%%%%%%%%%%%%%%%%%%%%%%%%%%%

\end{document}